\documentclass[
twocolumn,
superscriptaddress,
amsmath,amssymb,
prl,
]{revtex4-2}
\usepackage{graphicx}
\usepackage{hyperref}
\usepackage{dcolumn}
\usepackage{bm}
\usepackage{epstopdf}
\usepackage{romannum}
\usepackage{csquotes}
\usepackage[dvipsnames]{xcolor}
\usepackage{hyperref}

\begin{document}
	
	\preprint{EuFeAsP}
	
	\title{Multiple magnetic orders discovered in the superconducting state of EuFe$_{2}$(As$_{1-x}$P$_{x}$)$_{2}$}
	
	\author{Nan Zhou}
	\affiliation{School of Physics, Southeast University, Nanjing 211189, China}
	\affiliation{Institute for Solid State Physics (ISSP), The University of Tokyo, Kashiwa, Chiba 277-8581, Japan}
	\affiliation{Key Laboratory of Materials Physics, Institute of Solid State Physics, HFIPS, Chinese Academy of Sciences, Hefei 230031, China}
	
	\author{Yue Sun}
	\email{Corresponding author:sunyue@seu.edu.cn}
	\affiliation{School of Physics, Southeast University, Nanjing 211189, China}
	
	\author{Ivan S. Veshchunov}
	\affiliation{Department of Applied Physics, University of Tokyo, 7-3-1 Hongo, Bunkyo-ku, Tokyo 113-8656, Japan}

	\author{S. Kittaka}
	\affiliation{Department of Physics, Faculty of Science and Engineering, Chuo University, Tokyo 112-8551, Japan}

	\author{X. L. Shen}
	\affiliation{Institute for Solid State Physics (ISSP), The University of Tokyo, Kashiwa, Chiba 277-8581, Japan}
	
	\author{H. M. Ma}
	\affiliation{Institute for Solid State Physics (ISSP), The University of Tokyo, Kashiwa, Chiba 277-8581, Japan}
	
	\author{W. Wei}
	\affiliation{School of Physics, Southeast University, Nanjing 211189, China}

	\author{Y. Q. Pan}
	\affiliation{School of Physics, Southeast University, Nanjing 211189, China}
	\affiliation{Institute for Solid State Physics (ISSP), The University of Tokyo, Kashiwa, Chiba 277-8581, Japan}
	\affiliation{Key Laboratory of Materials Physics, Institute of Solid State Physics, HFIPS, Chinese Academy of Sciences, Hefei 230031, China}
	
	\author{M. Cheng}
	\affiliation{Key Laboratory of Materials Physics, Institute of Solid State Physics, HFIPS, Chinese Academy of Sciences, Hefei 230031, China}
	
	\author{Y. F. Zhang}
	\affiliation{School of Physics, Southeast University, Nanjing 211189, China}
	\affiliation{Institute for Solid State Physics (ISSP), The University of Tokyo, Kashiwa, Chiba 277-8581, Japan}
	
	\author{Y. Kono}
	\affiliation{Department of Physics, Faculty of Science and Engineering, Chuo University, Tokyo 112-8551, Japan}

	\author{Yuping Sun}
	\affiliation{Key Laboratory of Materials Physics, Institute of Solid State Physics, HFIPS, Chinese Academy of Sciences, Hefei 230031, China}
	\affiliation{High Magnetic Field Laboratory, HFIPS, Chinese Academy of Sciences, Hefei 230031, China}
	\affiliation{Collaborative Innovation Centre of Advanced Microstructures, Nanjing University, Nanjing 210093, China}
	
	\author{T. Tamegai}
	\affiliation{Department of Applied Physics, University of Tokyo, 7-3-1 Hongo, Bunkyo-ku, Tokyo 113-8656, Japan}

	\author{Xuan Luo}
	\email{Corresponding author:xluo@issp.ac.cn}
	\affiliation{Key Laboratory of Materials Physics, Institute of Solid State Physics, HFIPS, Chinese Academy of Sciences, Hefei 230031, China}
	
	\author{Zhixiang Shi}
	\email{Corresponding author:zxshi@seu.edu.cn}
	\affiliation{School of Physics, Southeast University, Nanjing 211189, China}

	\author{Toshiro Sakakibara}
	\affiliation{Institute for Solid State Physics (ISSP), The University of Tokyo, Kashiwa, Chiba 277-8581, Japan}

\date{\today}

\begin{abstract}

The interplay between superconductivity and magnetism is an important subject in condensed matter physics. EuFe$_{2}$As$_{2}$-based iron pnictides could offer an interesting plateau to study their relationship that has attracted considerable attention. So far, two magnetic phase transitions were observed in EuFe$_{2}$As$_{2}$-based crystal, which were deemed to originate from the itinerant Fe moments ($\sim$ 190 K) and the localized Eu$^{2+}$ moments ($\sim$ 19 K), respectively. Here, we systematically studied the heat capacity for the EuFe$_{2}$(As$_{1-x}$P$_{x}$)$_{2}$ crystals with \textit{x} = 0.21 (optimally doped) and \textit{x} = 0.29 (overdoped). We have found two new magnetic orders in the superconducting state (ranging from 0.4 to 1.2 K) in the optimally doped crystal. As more P was introduced into the As site, one of the magnetic orders becomes absent in the overdoped crystal. Additionally, we observed strong field and orientation dependence in heat capacity. The present findings in EuFe$_{2}$(As$_{1-x}$P$_{x}$)$_{2}$ have detected the new low-temperature magnetic orders, which may originate from the localized Eu$^{2+}$ spins order or the spin reorientation.

\end{abstract}
\pacs{\\
E-mail: sunyue@seu.edu.cn, xluo@issp.ac.cn, zxshi@seu.edu.cn}
\maketitle

The coexistence of magnetism and superconductivity (SC) has been of great interest because of the two phenomena usually are antagonistic and represent the competing nature \cite{clogston1962upper, saxena2000superconductivity, aoki2001coexistence, fertig1977destruction, ishikawa1977destruction, felner1997coexistence, bernhard1999coexistence, aoki2012ferromagnetism, pachmayr2015coexistence, saha2009superconducting, luo2011interplay, cao2010self, ren2009superconductivity}. Generally, SC is easily destroyed by ferromagnetism (FM) \cite{clogston1962upper}, and two states are always believed to originate from different electrons of different elements. However, triggered by certain conditions, some materials could hold both FM and SC states \cite{saxena2000superconductivity, aoki2001coexistence, fertig1977destruction, ishikawa1977destruction, felner1997coexistence, bernhard1999coexistence, aoki2012ferromagnetism, pachmayr2015coexistence, saha2009superconducting, luo2011interplay, cao2010self, ren2009superconductivity}, and some studies even found that SC and FM could come from the same type of electrons when SC emerges under the ferromagnetic background (\textit{T}$_{\rm{c}}$ $<$ \textit{T}$_{\rm{FM}}$) \cite{saxena2000superconductivity, aoki2001coexistence}. Earlier experiments demonstrated the coexistence of SC and FM in compounds such as ErRh$_{4}$B$_{4}$ \cite{fertig1977destruction}, Ho$_{1.2}$Mo$_{6}$S$_{8}$ \cite{ishikawa1977destruction}, and some high-\textit{T}$_{\rm{c}}$ rutheno-cuprates \cite{felner1997coexistence, bernhard1999coexistence}. Subsequently, this phenomenon was also observed in some uranium-based heavy-fermion compounds \cite{saxena2000superconductivity, aoki2001coexistence, aoki2012ferromagnetism} and iron-based superconductors \cite{pachmayr2015coexistence, saha2009superconducting, luo2011interplay, cao2010self}. Most of these compounds possess weak FM with small ordered moment. Recently, SC coexists with a strong ferromagnetic order were observed in EuFe$_{2}$As$_{2}$-family iron pnictides \cite{ren2009superconductivity}. The distinctive interaction between rare-earth Eu and FeAs layers has attracted consideration attention in condensed matter physics \cite{zapf2011varying, tokiwa2012unique, nandi2014magnetic}. The above studies suggest that the coexistence states between SC and FM will show intriguing physics. In order to fully understand the studied issue, future efforts on exploring materials with magnetism and SC or detecting the new magnetic orders under the superconducting state will be important.

\begin{figure*}
\includegraphics[width=42.5pc]{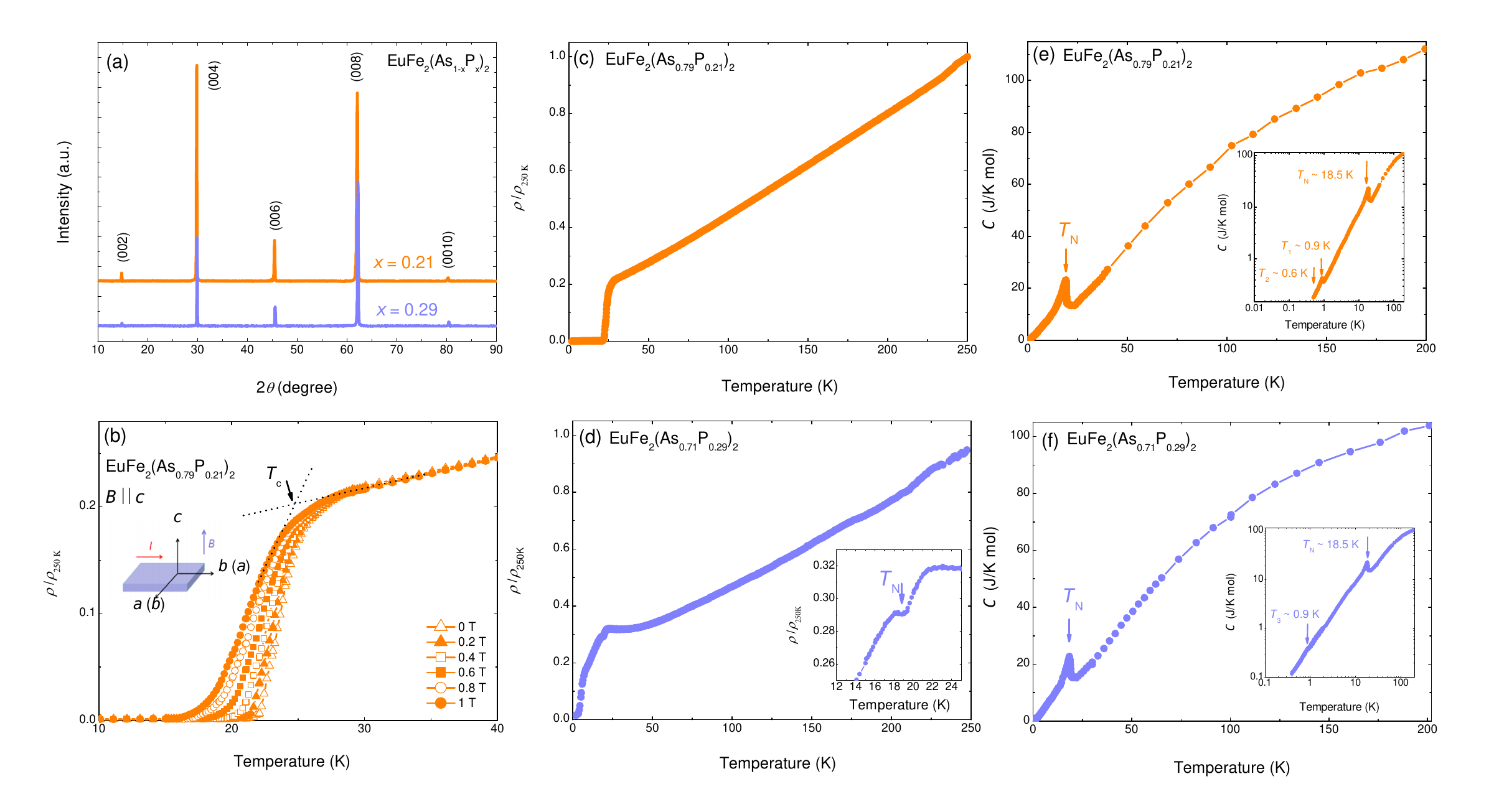}
\begin{center}
\caption{\label{S1} (a) Room temperature XRD patterns for the two P-doped crystals studied in this paper. (b) Temperature dependence of the resistivity for EuFe$_{2}$(As$_{0.79}$P$_{0.21}$)$_{2}$ with \textit{B} $\parallel$ \textit{c} under various fields from 0 to 1 T. The inset show the configuration of the resistivity. The direction of the electrical current along the \textit{a} or \textit{b} axis in the \textit{ab} plane. Temperature dependence of the resistivity at zero field for (c) EuFe$_{2}$(As$_{0.79}$P$_{0.21}$)$_{2}$ and (d) EuFe$_{2}$(As$_{0.71}$P$_{0.29}$)$_{2}$, respectively. The inset of (d) is a blow-up of the magnetic transition. Note that all curves are renormalized to the $\rho$$_{250 K}$ values. Temperature dependence of specific heat for (e) EuFe$_{2}$(As$_{0.79}$P$_{0.21}$)$_{2}$ and (f) EuFe$_{2}$(As$_{0.71}$P$_{0.29}$)$_{2}$, respectively. The inset shows the log-scale plot of the temperature dependent heat capacity.}\label{}
\end{center}
\end{figure*}

The target material of this study is the prototypical 122-family iron pnictide EuFe$_{2}$(As$_{1-x}$P$_{x}$)$_{2}$ \cite{ren2009superconductivity, zapf2013eufe, stolyarov2018domain, ghimire2021effect}. In the \textit{parent} EuFe$_{2}$As$_{2}$ compound, a combined transition of structural and spin-density-wave (SDW) orders of Fe magnetic moments were observed at \textit{T} $\sim$ 190 K. The rare earth Eu$^{2+}$ also develops a magnetic order at \textit{T} $\sim$ 19 K with a ferromagnetic alignment of the moments along \textit{a} axis and antiferromagnetic coupling along \textit{c} axis \cite{jeevan2008electrical, herrero2009magnetic, xiao2009magnetic, jiang2009metamagnetic}. By doping P in EuFe$_{2}$As$_{2}$, the structural transition and SDW order are gradually suppressed and the SC emerges \cite{jeevan2011interplay, xu2014electronic}. Intriguingly, the magnetic order of Eu$^{2+}$ moments changes from the antiferromagnetic coupled at low P concentrations to a ferromagnetic state on the overdoped range along the \textit{c} axis \cite{zapf2011varying, tokiwa2012unique, nandi2014magnetic}.

In this letter, we present a systematic investigation of the heat capacity in P-doped EuFe$_{2}$As$_{2}$ single crystals. Two new magnetic orders were observed at very low temperatures (0.4 to 1.2 K) in EuFe$_{2}$(As$_{0.79}$P$_{0.21}$)$_{2}$. As the P content is increased (overdoped sample, EuFe$_{2}$(As$_{0.71}$P$_{0.29}$)$_{2}$), one of the magnetic orders disappears. Moreover, both the P-doped EuFe$_{2}$As$_{2}$ samples display anisotropic heat capacity with strongly field and orientation dependence. We attribute the low-temperature magnetic orders to the localized Eu$^{2+}$ spins order or the spin reorientation, which will provide new insights to understand this system.

\begin{figure*}
\includegraphics[width=42.5pc]{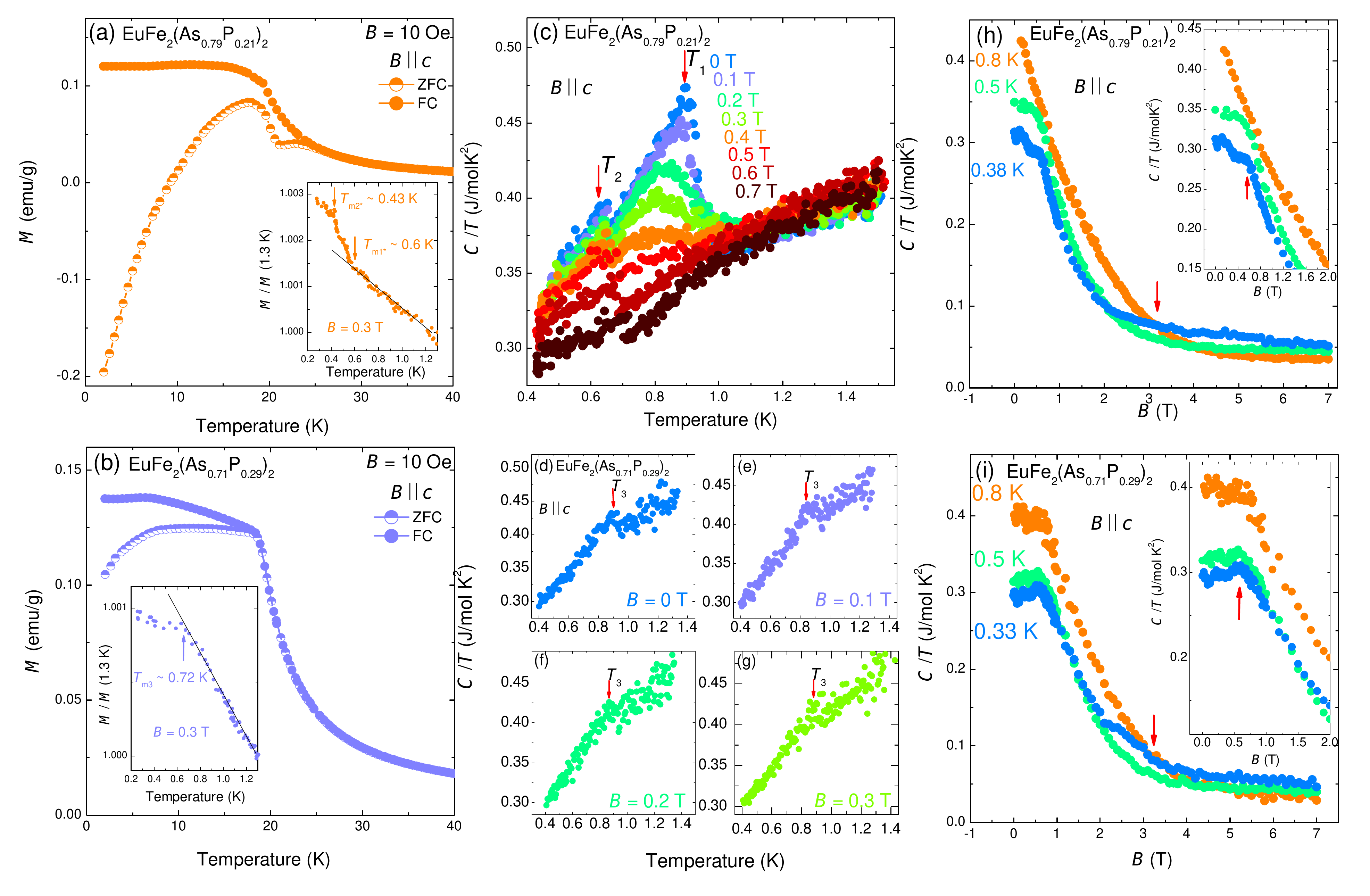}
\begin{center}
\caption{\label{fig1} (a) The magnetizations with ZFC (half solid circle) and FC (solid circle) modes for (a) EuFe$_{2}$(As$_{0.79}$P$_{0.21}$)$_{2}$ and (b) EuFe$_{2}$(As$_{0.71}$P$_{0.29}$)$_{2}$, respectively. The inset shows the results of temperature dependent dc magnetization under 0.3 T at very low temperature range, which were measured by using the Capacitance-Faraday method. These characteristic points (\textit{T}$_{m1*}$, \textit{T}$_{m2*}$, and \textit{T}$_{m3}$) were obtained. (c) Temperature dependences of \textit{C}(\textit{T})/\textit{T} for EuFe$_{2}$(As$_{0.79}$P$_{0.21}$)$_{2}$ crystal (\textit{B} $\parallel$ \textit{c}) at various fields. (d)--(g) Temperature dependences of \textit{C}(\textit{T})/\textit{T} for EuFe$_{2}$(As$_{0.71}$P$_{0.29}$)$_{2}$ crystal (\textit{B} $\parallel$ \textit{c}) at various fields. Magnetic field dependences of \textit{C}(\textit{B})/\textit{T} for (h) EuFe$_{2}$(As$_{0.79}$P$_{0.21}$)$_{2}$ and (i) EuFe$_{2}$(As$_{0.71}$P$_{0.29}$)$_{2}$, respectively. The inset blows up the low temperature region.}
\end{center}
\end{figure*}

The single crystals of EuFe$_{2}$(As$_{0.79}$P$_{0.21}$)$_{2}$ and EuFe$_{2}$(As$_{0.71}$P$_{0.29}$)$_{2}$ studied here were synthesized using the self-flux method \cite{vinnikov2019direct, veshchunov2017visualization}.  The actual chemical composition was determined via energy dispersive x-ray spectroscopy (EDX). The structure was characterized by x-ray diffraction (XRD) using a Rigaku diffractometer with Cu-\textit{K$\alpha$} radiation. The directions of the crystal axes were determined by single-crystal XRD. The resistivity and high-temperature heat capacity were measured by using a physical property measurement system (PPMS, Quantum Design). The magnetization was measured by using a commercial superconducting quantum interference device magnetometer (MPMS-XL5, Quantum Design). The field-orientation dependence of the specific heat was measured in an 8 T split-pair superconducting magnet with a $^{3}$He refrigerator. The refrigerator can be continuously rotated by a motor on top of the Dewar with an angular resolution better than 0.01$^{\circ}$ \cite{sun2017gap, sun2018disorder, sun2019quasiparticle}. The definitions of the in-plane (azimuthal) angle $\phi$ and the out-of-plane (polar) angle $\theta$ are shown in the inset of Fig. 2(a).

\begin{figure*}
\includegraphics[width=42.5pc]{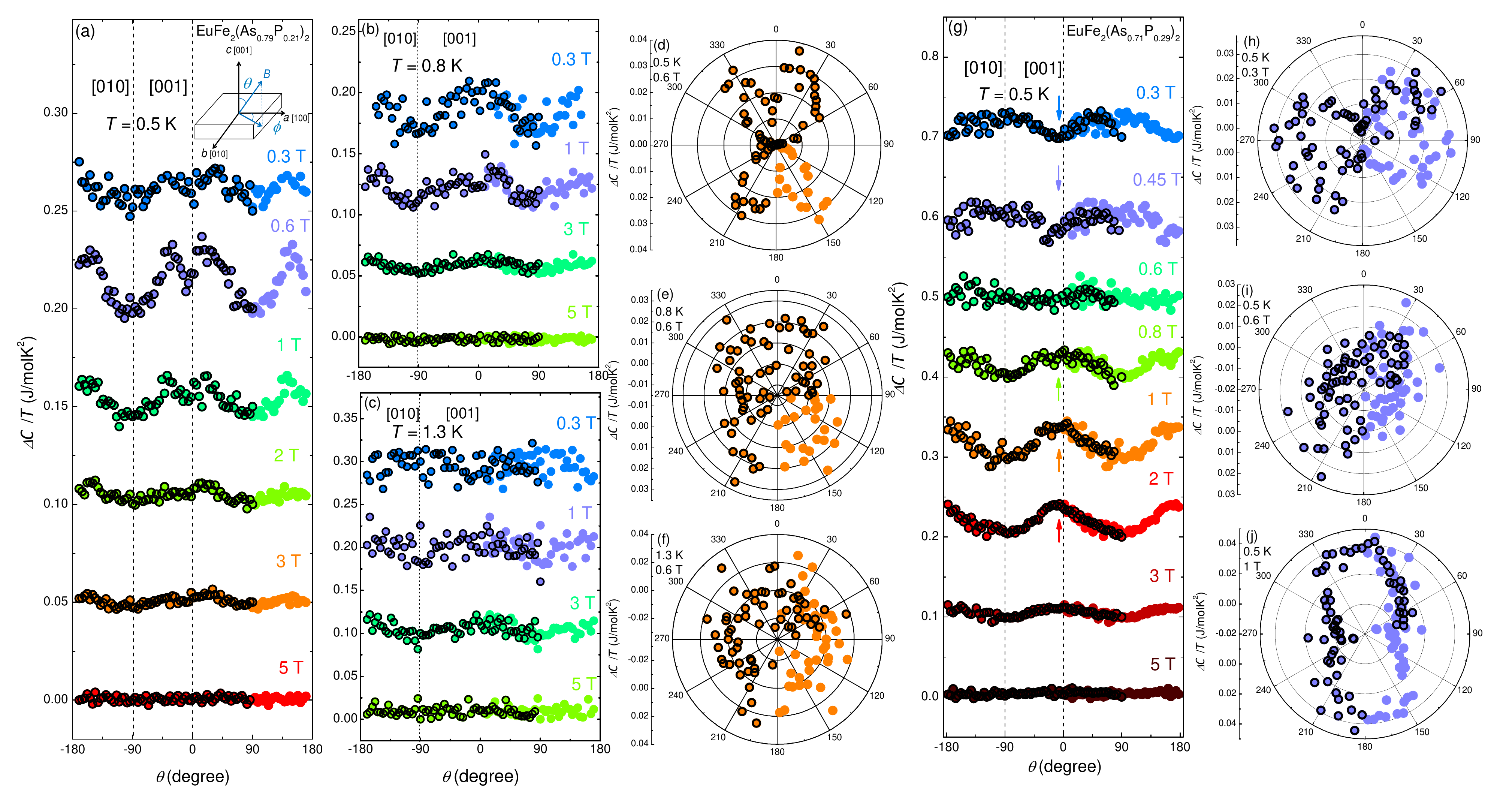}
\begin{center}
\caption{\label{fig2} (a) Azimuthal angle dependence of the specific heat \textit{$\Delta$}\textit{C}(\textit{$\theta$})/\textit{T} measured under various fields at 0.5 K for EuFe$_{2}$(As$_{0.79}$P$_{0.21}$)$_{2}$ crystal. \textit{$\Delta$}\textit{C}(\textit{$\theta$})/\textit{T} is defined as \textit{C}(\textit{$\theta$})/\textit{T}-\textit{C}(90$^{\circ}$)/\textit{T}, and each subsequent curve is shifted vertically by 0.05 J/molK$^{2}$. Symbols with black outlines are measured data (ranging from -180 $^{\circ}$ to 90 $^{\circ}$), and those without are mirrored points to show the symmetry (ranging from 0 $^{\circ}$ to 180 $^{\circ}$). Definitions of azimuthal (\textit{$\phi$}) and polar (\textit{$\theta$}) angles with respect to the crystal axes, as shown in the inset. (b) \textit{$\Delta$}\textit{C}(\textit{$\theta$})/\textit{T} measured at 0.8 K under various fields. Each subsequent curve is shifted by 0.05 J/molK$^{2}$. (c) \textit{$\Delta$}\textit{C}(\textit{$\theta$})/\textit{T} measured at 1.3 K under various fields. Each subsequent curve is shifted by 0.1 J/molK$^{2}$. (d)--(f) Polar plot of the out of plane \textit{$\Delta$}\textit{C}(\textit{$\theta$})/\textit{T} for EuFe$_{2}$(As$_{0.79}$P$_{0.21}$)$_{2}$ crystal under 0.6 T at 0.5, 0.8, and 1.3 K, respectively. (g) Azimuthal angle dependence of the specific heat \textit{$\Delta$}\textit{C}(\textit{$\theta$})/\textit{T} measured at 0.5 K under various fields. Each subsequent curve is shifted vertically by 0.1 J/mol K$^{2}$. (h)--(j) Polar plot of the out of plane \textit{$\Delta$}\textit{C}(\textit{$\theta$})/\textit{T} for EuFe$_{2}$(As$_{0.71}$P$_{0.29}$)$_{2}$ crystal at 0.5 K under 0.3, 0.6, and 1 T, respectively.}
\end{center}
\end{figure*}

The XRD patterns of P-doped EuFe$_{2}$As$_{2}$ crystals are shown in Fig. 1(a). Only the (00\textit{$\ell$}) peaks are observed, which can be well indexed based on a tetragonal struture with the \textit{I}4/\textit{mmm} space group. The positions of the (00\textit{$\ell$}) peaks were found to be shifted to higher angles with increasing the P content, which indicates the shrinkage in \textit{c}-lattice parameters (see Fig. 1(a)). The \textit{c}-axis parameters are evaluated as $\sim$ (11.913 $\pm$ 0.01) ${\AA}$ and (11.883 $\pm$ 0.01) ${\AA}$ for the optimally doped and overdoped crystals, respectively. In order to evaluate the chemical composition for our crystals, some \textit{c}-axis lattice parameters of the previous results are also incorporated for comparison \cite{cao2011superconductivity}. We found that they could match each other well. Temperature dependence of resistivity are shown in Figs. 1(b)-(d). Figure 1(b) shows the low temperature superconducting transitions below 40 K for the optimally doped crystal. The inset show the configuration of the resistivity. The direction of the electrical current along the \textit{a} or \textit{b} axis in the \textit{ab} plane. \textit{T}$_{\rm{c}}$ is determined by the criterion of the onset of the superconducting transition. The nonzero resistivity is observed in the overdoped crystal (see Fig. 1(d)). In actually, these non-SC crystals under the overdoped side usually show a relativity poor quality, which have been reported in the previous study \cite{xu2014electronic}. In our case, a small amount of elements vacancies have been observed in our overdoped crystal. Meanwhile, an incipient dip related to the Eu$^{2+}$ magnetic transition occurs around 20 K, and a quick drop below $\sim$ 7 K is observed, which is similar to the previous results \cite{xu2014electronic}. In order to confirm these behavior, we also carefully check the heat capacity and magnetic susceptibility, which have been considered as the thermodynamic measurements. We don't find any anomalous behavior. After comprehensive consideration, we think the sample inhomogeneity or a strong internal field induced by Eu$^{2+}$ ferromagnetic ordering on the overdoped side could be a possible scenario for understanding these phenomenons \cite{xu2014electronic}. The temperature dependence of zero-field-cooled (ZFC) and field-cooled (FC) dc magnetic susceptibility under various fields are shown in Figs. S1, 2(a), and 2(b). The supercondcuting diamagnetic signal has been observed under 10 Oe in the optimal doped crystal (see Fig. 2(a)), while the ferromagnetic order of the Eu$^{2+}$ moments dominant the magnetic susceptibility in overdoped crystal (see Fig. 2(b)). Figures 1(e) and 1(f) show the results of heat capacity measurements. The heat capacity anomalies could be clearly observed below 30 K in both the two P-doped samples. The $\lambda$-shape peak (\textit{T}$_{\rm{N}}$ $\sim$ 18.5 K) is attributed to a ferromagnetic nature from the Eu$^{2+}$ moments. These results confirm the basic characteristic for the present P-doped crystals, which are consistent to the previous studies \cite{ren2009superconductivity, jeevan2011interplay, xu2014electronic}. Furthermore, at lower temperature region, two prominent heat capacity jumps at \textit{T}$_{1}$ and \textit{T}$_{2}$ are observed in the optimally doped crystals (see the inset of Fig. 1(e)), while only one heat capacity jump at \textit{T}$_{3}$ is observed in the overdoped crystal (see the inset of Fig. 1(f)).


\begin{figure*}
\includegraphics[width=42.5pc]{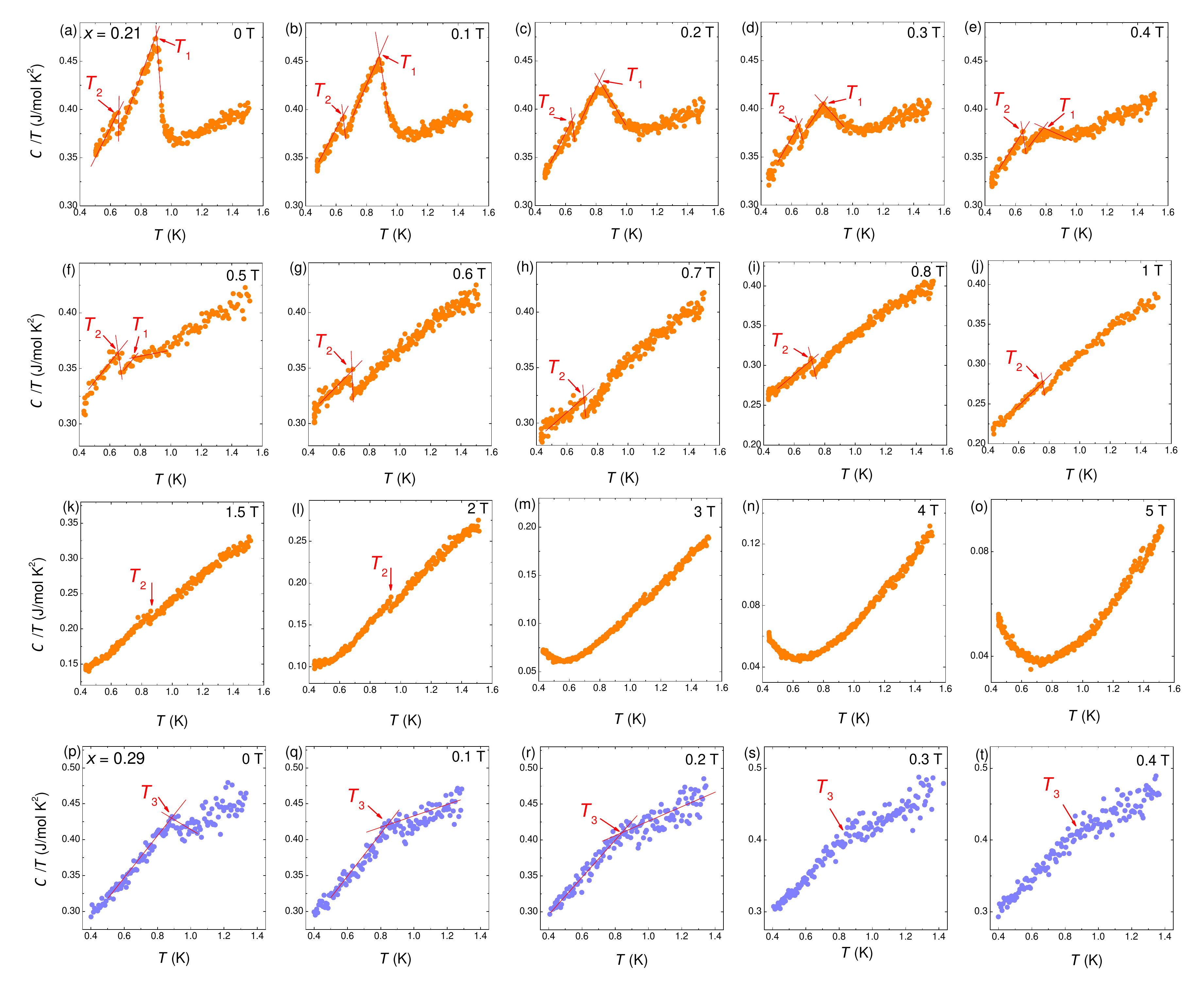}
\begin{center}
\caption{\label{fig3} Field-temperature-resolved specific heat for (a)--(o) the optimally doped crystals and (p)--(t) the overdoped crystals. These characteristic points (\textit{T}$_{1}$, \textit{T}$_{2}$, and \textit{T}$_{3}$) were determined from the intersection of two extrapolated red lines.}
\end{center}
\end{figure*}

\begin{figure*}
\includegraphics[width=42.5pc]{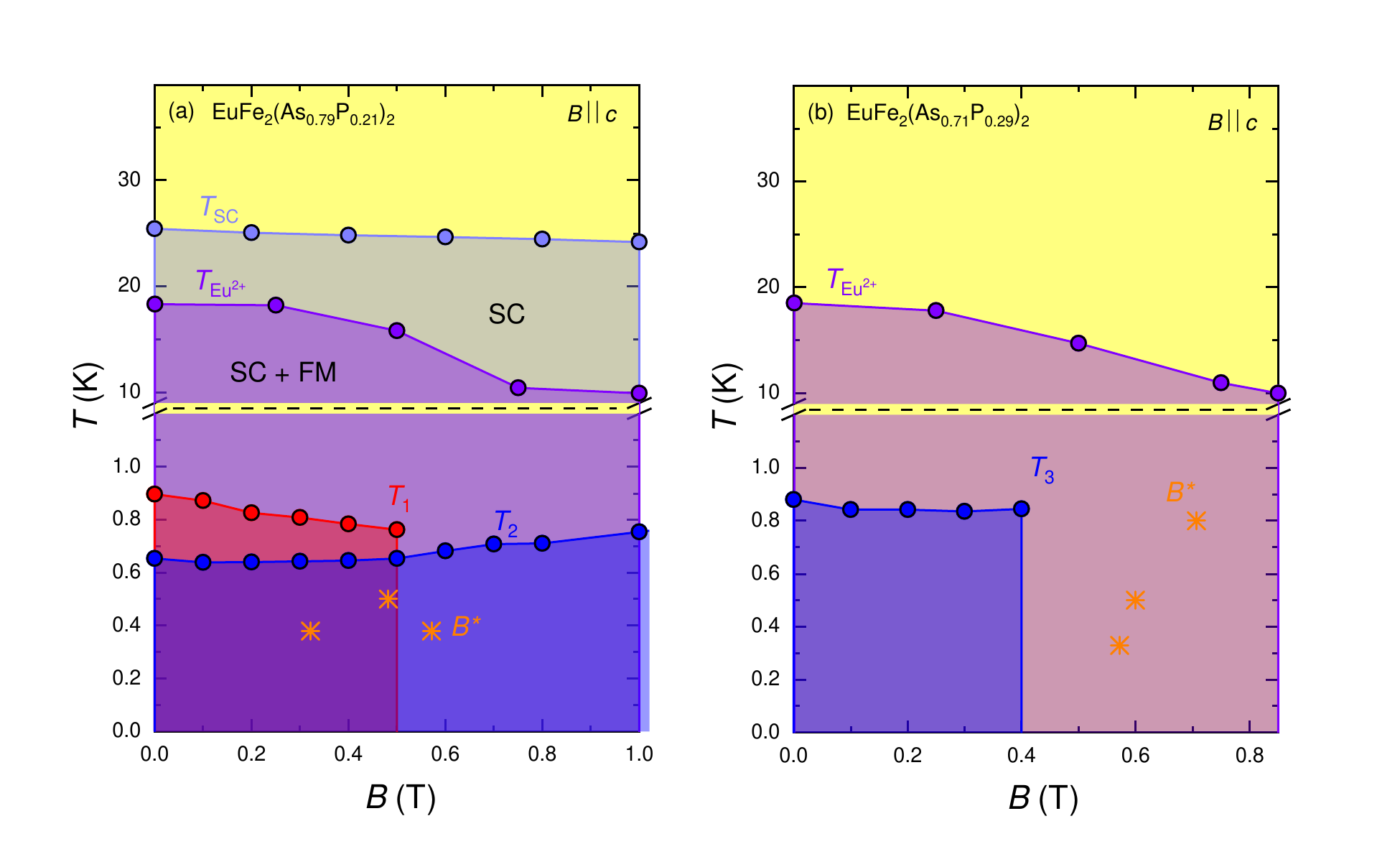}
\begin{center}
\caption{\label{fig3} Magnetic-temperature phase diagram determined from the field angle-resolved specific heat measurements for (a) EuFe$_{2}$(As$_{0.79}$P$_{0.21}$)$_{2}$ and (b) EuFe$_{2}$(As$_{0.71}$P$_{0.29}$)$_{2}$, respectively. The applied field is parallel to the crystallographic \textit{c}-axis. \textit{T}$_{SC}$ (light blue) denotes the superconducting transition temperature. \textit{T}$_{Eu^{2+}}$ (purple) denotes the magnetic transition of the Eu$^{2+}$ ions. \textit{T}$_{1}$ (red) and \textit{T}$_{2}$ (blue) correspond to the peak of the phase transitions for EuFe$_{2}$(As$_{0.79}$P$_{0.21}$)$_{2}$ crystal. \textit{T}$_{3}$ (blue) corresponds to the peak of the phase transitions for EuFe$_{2}$(As$_{0.71}$P$_{0.29}$)$_{2}$ crystal. \textit{B}$^{*}$ were extracted from the field dependent heat capacity. The more detail of the determinations about these characteristic points are shown in Fig. 3.}
\end{center}
\end{figure*}

In order to obtain more information about these new low-temperature phases, we systematically investigated their heat capacity behavior under various fields. Figure 2(c) shows the \textit{C}(\textit{T})/\textit{T} curves of EuFe$_{2}$(As$_{0.79}$P$_{0.21}$)$_{2}$ measured at various fields up to 0.7 T applied along \textit{c} axis. Two prominent jumps occur at \textit{T}$_{1}$(0 T) $\sim$ 0.9 K and \textit{T}$_{2}$(0 T) $\sim$ 0.6 K. 
With increasing fields, the first transition \textit{T}$_{1}$ shifts to lower temperatures, and the second transition \textit{T}$_{2}$ shifts to the high temperature region. At around 0.5 T, two transitions become very close to each other (see Fig. 2(c)). As magnetic field further increases, \textit{T}$_{1}$ is nearly invisible, while \textit{T}$_{2}$ continually shifts to higher temperatures (see Fig. S2(a) \cite{supplement}). Meanwhile, we noticed that a significant upturn under higher fields at very low temperatures for both the two doped crystals, which are attributed to the Schottky behaviour of the nuclear contributions. Moreover, the field dependences of the heat capacity are also performed as shown in Fig. 2(h). The \textit{C}(\textit{B})/\textit{T} exhibits a rapidly increase below a characteristic field of approximately 3 T. At lower fields, an obvious heat capacity anomaly is observed at 0.38 K and 0.5 K (below \textit{T}$_{2}$), while they are absent at 0.8 K, indicating that \textit{T}$_{1}$ and \textit{T}$_{2}$ phases correspond to different magnetic response in EuFe$_{2}$(As$_{0.79}$P$_{0.21}$)$_{2}$ crystal (see the inset of Fig. 2(h)). On the other hand, there is only one heat capacity jump observed at \textit{T}$_{3}$(0 T) $\sim$ 0.9 K in the overdoped EuFe$_{2}$(As$_{0.71}$P$_{0.29}$)$_{2}$ crystal (see Figs. 2(d)--(g)). The specific heat kink gradually disappears when increases the magnetic field (see Fig. S2(b) \cite{supplement}). The \textit{C}(\textit{B})/\textit{T} displays a similar behaviour with the \textit{T}$_{2}$ phase in optimal doped crystal (see Fig. 2(i)), indicating that they may originate from the same source.

In order to learn more about the anomalous jump in heat capacity, we systematically investigated the field angle-resolved specific heat for the two P-doped crystals. The field direction is schematically shown in the inset of Fig. 3(a). To obtain a clear view of the symmetry, each subsequent curve is shifted vertically by a fixed value. Symbols with black outlines are measured data, and those without are mirrored points to show the symmetry. Definitions of azimuthal ($\varphi$) and polar ($\theta$) angles with respect to the crystal axes are shown in the inset of Fig. 3(a). The out-of-plane heat capacity results are presented in detail in Figs. 3(a)-(f) for the optimally doped crystal. At 1.3 K, there is almost no significantly angular dependence in \textit{C}($\theta$)/\textit{T} under 0.3 T and 5 T. A slight twofold symmetry observed when the magnetic field is fixed at 1 T and 3 T (see Fig. 3(c)). The weak twofold signal may originate from the inherent tetragonal symmetry or the magnetic order ($\sim$ 19 K). As the sample is cooled to 0.8 K (\textit{T}$_{2}$ $<$ \textit{T} $<$ \textit{T}$_{1}$), the heat capacity data exhibits a clear twofold symmetry at 0.3 T, 1 T, and 3 T (see Fig. 3(b)). Subsequently, as the temperature continues to decrease below \textit{T}$_{2}$ phase, the twofold symmetric curve evolves to a butterfly-pattern with fourfold symmetry at 0.5 K(see Fig. 3(a)). The evolution of the symmetry at various temperatures under 0.6 T were shown in Fig. S3(a) \cite{supplement}. It can be visualized more clearly in the polar plots presented in Figs. 3(d)--(f).
Meanwhile, we also noticed that the anisotropy of the heat capacity is absent at any temperature under a higher field ($\sim$ 5 T). To obtain more information about the low-temperature anomalous transitions, we perform the angular dependent heat capacity with the field rotation in the FeAs plane (see Fig. S4 \cite{supplement}). Unlike the behaviour of the out-of-plane heat capacity, the in-plane heat capacity shows a relatively weak oscillation. The clear signal could only be observed when the temperature fixed at 0.5 K below \textit{T}$_{2}$. At 0.8 K (\textit{T}$_{2}$ $<$ \textit{T} $<$ \textit{T}$_{1}$), it is difficult to distinguish the symmetry on the basis of the current results. The scattered data were displayed at 1.3 K (\textit{T} $>$ \textit{T}$_{1}$), indicating the weak anisotropic heat capacity.

In analogy to EuFe$_{2}$(As$_{0.79}$P$_{0.21}$)$_{2}$, the field and angular dependence of the heat capacity have also been performed at various temperatures for the overdoped sample, as shown in Figs. 3(g)--(j), Figs. S3(b)--(e), and Fig. S5 \cite{supplement}. The out of plane heat capacity \textit{C}(\textit{$\theta$})/\textit{T} exhibits twofold symmetry when the temperature is fixed at 0.5 K. Interestingly, it is found that the heat capacity waveform could be tuned by the field. The peak of the twofold symmetric curve is changed to the valley by increasing the magnetic field from 0.3 T to 1 T (see Fig. 3(g)). The absence of the anisotropy under 0.6 T could be understand that the value of the peak and valley cancel each other out, which is almost consistent with the anomalies of the \textit{C}(\textit{B})/\textit{T} data, as shown in the inset of the Fig. 2(i). In order to obtain more information about the reversed heat capacity signal, we performed the angular dependent heat capacity at various temperatures under 0.3 T, 0.6 T, and 1 T (see Fig. S5 \cite{supplement}), respectively. The three sets of fields correspond to different heat capacity waveforms. With increasing temperature, the heat capacity anisotropy measured under 0.3 T and 1 T gradually smear, while they keep isotropic under 0.6 T at all temperature points. A better view of the symmetry is displayed in Figs. 3(h)-(j) by the polar plots, which shows that the direction of the heat capacity symmetry changes apparently at 0.5 K between 0.3 T and 1 T. Furthermore, similar to the EuFe$_{2}$(As$_{0.79}$P$_{0.21}$)$_{2}$, the anisotropy is also absent at any temperature under a higher field ($\sim$ 5 T). Correspondingly, unlike the case of the optimally doped compound, we can observe stronger in-plane heat capacity in the overdoped crystal (see Fig. S3(b) \cite{supplement}). The reversed waveforms are also observed when the field rotates in the FeAs plane, which could be seen more clear in the polar plots presented in Figs. S3(c)--(e) \cite{supplement}.

Figure 5 shows the \textit{B}-\textit{T} phase diagram revealed from our field temperature-resolved specific heat measurements for the optimally doped and overdoped EuFe$_{2}$(As$_{1-x}$P$_{x}$)$_{2}$ crystals, respectively. The characteristic temperatures of the superconducting state (\textit{T}$_{\rm{SC}}$, light blue) and the localized Eu$^{2+}$ moments (\textit{T}$_{\rm{Eu^{2+}}}$, purple) could be obtained from the electrical transport and magnetization, as shown in Figs. 1 and 2. These characteristic points (\textit{T}$_{1}$, \textit{T}$_{2}$, and \textit{T}$_{3}$) are shown in Fig. 4, which were determined from the intersection of two extrapolated red lines. \textit{T}$_{1}$ and \textit{T}$_{2}$ correspond to the peak of these phase transitions for the optimally doped crystal, while the \textit{T}$_{3}$ is for the overdoped crystal. With increasing the fields, the \textit{T}$_{2}$ exhibit a quasi-linear behaviour. Interestingly, \textit{T}$_{2}$ and \textit{T}$_{3}$ show similar behaviour in two different crystals, indicating that they may have the same microscopic origin. However, the difference is that the \textit{T}$_{2}$ phase could survive under higher fields in the optimal doping, while the \textit{T}$_{3}$ phase could be only observed below $\sim$ 0.4 T in overdoped crystal. In our opinion, the main reason for this behavior is that it is difficult for distinguishing the anomalies heat capacity under higher fields. Future efforts on more sensitive transport or magnetic susceptibility measurements are hopefully to establish a completed phase diagram. Meanwhile, in order to reflect a more actual situation, we extracted the characteristic points \textit{B}$^{*}$ from the field dependent heat capacity data in Fig. 2(h) and 2(i), which have also been included in the current phase diagram.

A natural question raised by our study concerns the microscopic origin of these new phase transitions and their relationship with various orders in EuFe$_{2}$(As$_{1-x}$P$_{x}$)$_{2}$. We notice that the characteristic temperature \textit{T}$_{1}$ shifts to lower temperature and becomes broadened with increasing the fields, while \textit{T}$_{2}$ and \textit{T}$_{3}$ shift to higher temperature region. Usually, a ferromagnetic or antiferromagnetic nature could induce this kind of \textit{$\lambda$}-shape peak in heat capacity data, which could drive these transition to higher and lower temperatures when increasing the magnetic field \cite{ren2009superconductivity}. In order to confirm the possible scenario of the magnetic orders, we measured the dc magnetization down to $\sim$ 0.2 K by using the Capacitance-Faraday method (see the inset of Figs. 1(a) and 1(b), Fig. S6 \cite{supplement}). We found that anomalous magnetization happened in both the two selected P-doped crystals. Two prominent magnetic transitions are observed under 0.3 T in the optimally doped crystal (see the inset of Fig. 2(a)). As magnetic field further increases, only one magnetic transition survives under 1 T (see Fig. S6(a) \cite{supplement}). In the overdoped case, there is only one magnetic transition under both the 0.3 T and 1 T (see the inset of Fig. 2(b) and Fig. S6(b) \cite{supplement}). Subsequently, we carefully compared these characteristic temperature points between heat capacity and dc magnetization measurements. We noticed that one phase and multi-phases transitions show different situations. In one phase transition case, these characteristic temperatures could match each other in the optimally doped crystal under 1 T (\textit{T}$_{2}$ $\sim$ 0.76 K, \textit{T}$_{m2}$ $\sim$ 0.66 K) and the overdoped crystal under 0.3 T (\textit{T}$_{3}$ $\sim$ 0.84 K, \textit{T}$_{m3}$ $\sim$ 0.72 K). On the other hand, two transitions have been observed in optimally doped crystal under 0.3 T. Temperature dependent magnetization shows a rapid increase (\textit{T}$_{m1*}$ $\sim$ 0.6 K) and lower temperature plateau (\textit{T}$_{m2*}$ $\sim$ 0.43 K) at very low temperature region (see the inset in Fig. 2(a)), which are different with other magnetization behavior that only contain one phase transition (see Fig. S6 \cite{supplement} and the inset in Fig. 2(b)). Correspondingly, the heat capacity characteristic temperatures are \textit{T}$_{1}$ $\sim$ 0.81 K and \textit{T}$_{2}$ $\sim$ 0.65 K. Combining the above results, we found that the characteristic points on magnetization always lag behind the heat capacity critical points. The absolute temperature difference is $\sim$ 0.1 K for one phase situation, and $\sim$ 0.2 K for the multi-phases transitions. The multi-phase system exhibits a relatively poorer matching relationship. Through the above results, we also have a doubt that whether this kind of regular temperature shift for the multi-phases system may be related to the interaction of the multi-phases? Certainly, the issue still under controversy, and will need to be systematically verified in the future. Although some issues still under discussion, we think these present dc magnetization results remain could be a persuasive evidence support the scenario of magnetic origin.

In EuFe$_{2}$As$_{2}$-based system, Eu$^{2+}$ moments and Fe$^{2+}$ moments are two primary sources of magnetism. Usually, the itinerant Fe$^{2+}$ moments correspond to a high-temperature ($\sim$ 190 K) magnetic phase transition, while the localized Eu$^{2+}$ moments show a relatively complex situation. The magnetic moment of the Eu$^{2+}$ ions could generate an appreciate ferromagnetic or antiferromagnetic contribution, which have been observed in many previous studies \cite{zapf2011varying, raffius1993magnetic, tan2016transition, anand2015antiferromagnetism, sengupta2005magnetic, jin2019spiral}. As in ThCr$_{2}$Si$_{2}$-type compounds, it is generally known that Eu$^{2+}$ exhibits a magnetic order at $\sim$ 19 K in EuFe$_{2}$As$_{2}$-based materials \cite{zapf2011varying}. EuCo$_{2}$As$_{2}$ \cite{raffius1993magnetic, tan2016transition}, EuCu$_{2}$As$_{2}$ \cite{anand2015antiferromagnetism, sengupta2005magnetic}, and EuNi$_{2}$As$_{2}$ \cite{jin2019spiral, raffius1993magnetic} show magnetic order at $\sim$ 47, 17.5, and 14 K, respectively. Two magnetic transitions related to the Eu$^{2+}$ ions have been reported. The first transition at 7.1 K, while the second one at 5.9 K in EuPt$_{2}$As$_{2}$ \cite{zhang2017magnetic}. \textit{T}$_{N1}$ = 11.0 K and \textit{T}$_{N2}$ = 5.5 K in EuPd$_{2}$As$_{2}$ \cite{anand2014physical}. More interestingly, some studies found that when Eu is substituted by Ba, Sr, and K, the magnetic transitions are absent in BaCo$_{2}$As$_{2}$ \cite{sangeetha2018enhanced}, SrPd$_{2}$As$_{2}$ \cite{anand2014physical}, and Eu$_{1-x}$K$_{x}$Fe$_{2}$As$_{2}$ \cite{jeevan2010superconductivity}, respectively. The above experimental evidences show that both of the characteristic temperature and the number of the magnetic transitions for Eu$^{2+}$ moments could be different in different systems. Thus, we realize that the localized Eu$^{2+}$ moments may play an important role for these new low-temperature phases. On the other hand, we also attempt to consider the possible scenario about the itinerant Fe$^{2+}$ moments. In the overdoped crystal, one of the new low-temperature magnetic orders and the high-temperature magnetic transitions ($\sim$ 190 K) disappearing together, which seem to mean that the two magnetic transitions share the same origin with Fe$^{2+}$ moments. However, in our opinion, there is still a lack of crucial evidence to confirm this issue. Meanwhile, we also check the heat capacity behavior in some similar materials that only containing Fe$^{2+}$ moments, such as FeSe \cite{sun2018disorder} and BaFe$_{2}$As$_{2}$ \cite{rotundu2010heat} crystals. Apart from the SC or the high-temperature magnetic transition, we don't find any anomalous heat capacity signal at low-temperature region. We think that future efforts on the temperature dependent heat capacity in doped sample (Eu site and Fe site) are hopefully to clarify this intrinsic origin. Finally, in order to give qualitative discussions on the released entropy base on the scenario of magnetic origin, we try to obtain the magnetic contribution to the heat capacity \textit{C}$_{mag}$(\textit{T}). The magnetic contribution to the heat capacity \textit{C}$_{mag}$(\textit{T}) is estimated from the zero-field \textit{C}(\textit{T}) data by subtracting the lattice contribution. The \textit{S}$_{mag}$(\textit{T}) was then determined by integrating the \textit{C}$_{mag}$(\textit{T})/\textit{T} versus $\textit{T}$ data. You can find more detailed information in Fig. S7 \cite{supplement}. For the optimally doped crystal, the \textit{S}$_{mag}$ attains a value of 14.3 J/mol K at \textit{T}$_{Eu^{2+}}$ which is 82.7\% of the expected high-\textit{T} limit \textit{R}ln(2\textit{S} + 1) = \textit{R}ln8 = 17.3 J/mol K for \textit{S} = 7/2. For the overdoped crystal, the value of 16.37 J/mol K at \textit{T}$_{Eu^{2+}}$ which is 94.62\%. Our results are almost consistent with the previous results, which the magnetic entropy is 16.5 J/mol K and 95\% for Eu$^{2+}$ \cite{ren2009superconductivity}. At low-temperature region, the \textit{C}$_{mag}$ jump have been observed, while the magnetic entropy \textit{S}$_{mag}$(\textit{T}) show a weak changed in both of the doped crystals. The similar behavior have been observed in EuPd$_{2}$As$_{2}$ crystal \cite{anand2014physical}, which suggest that the low temperature transitions is due to a spin reorientation. For these combined reasons, we tend to think that these new low-temperature magnetic orders have their origins related to the localized Eu$^{2+}$ spin order or their reorientation at the current stage.

As the above discussions, we attribute these new phases to the scenario of the localized Eu$^{2+}$ moments or the spin reorientation. Therefore, it's hard to ignore the research topic on the coexistence of SC and magnetism in the current work. Although, the issue has been discussed for a long time that focus on the magnetic transitions (10--30 K) \cite{zapf2011varying, tokiwa2012unique, nandi2014magnetic}. In analogy to these past discussions on the Eu$^{2+}$ order ($\sim$ 19 K), we think the current discovery could provide one more new opportunity. Therefore, on the basis of our results, some related research topics could be systematically studied in the future. For example, the magnetic structure, the more detailed evolution about how the new ordering impact on superconductivity, and whether there is spontaneous vortex phase or the reentrance effect occurred at the very low temperature region. In our opinion, the above relevantly issues are important not only for these new phases, but also for understanding the SC in iron-based superconductors, which will give a new insight to understand the EuFe$_{2}$As$_{2}$-based system.

For these anisotropic heat capacity, the optimally doped sample show a four-fold symmetry, while a significant two-fold symmetry behavior observed in non-SC overdoped crystal. Meanwhile, you can observe that there is almost no heat capacity oscillations when the temperature above the critical points for both the doped crystals. These results could confirm that the low-temperature heat capacity jump have a close relationship with the anisotropic heat capacity. However, in our opinion, the anisotropic superconducting gap is always excited at very low temperature range, which also could be a possible scenario for understanding the changed heat capacity symmetry in optimally doped crystal. Furthermore, it's also quite possible that there are strong field and orientation dependence in heat capacity at higher temperature (around 10--30 K), which could provide some useful information for understanding the Eu$^{2+}$ magnetic transition ($\sim$ 19 K). However, the present measurements of angle resolved specific heat (electronic specific heat) at high temperatures is influenced easily because large phononic and the background noise contributions to the total specific heat signal. Some relevant experimental details need to be repeatedly confirmed. Therefore, based on the current situation, it is very difficult to distinguish the angle dependence part. Future efforts on how to extract the intrinsic electronic specific heat are hopefully to solve this issue. At last, we try to discuss some possible situations on the external origins such as the impurity phase. In our case, the low-temperature heat capacity anomalies have been observed in different doping content crystals. Meanwhile, comparing with the fundamental data in previous reports, we did not find any significant differences to support the impurity scenario. Certainly, for some unconsidered situations, these issues are worthy of exploration in future.



To conclude, we systematically studied the heat capacity of the phosphorus doped EuFe$_{2}$As$_{2}$ crystals. New magnetic orders were observed at very low temperatures (range from 0.4 to 1.2 K) in EuFe$_{2}$(As$_{1-x}$P$_{x}$)$_{2}$ (\textit{x} = 0.21, 0.29) crystals, which also display anisotropic heat capacity with strongly field and orientation dependent. We attribute these new magnetic orders to the localized Eu$^{2+}$ spins order or the spin reorientation, which may provide new clues to understand the coexistence of magnetism and SC in EuFe$_{2}$(As$_{1-x}$P$_{x}$)$_{2}$.

The authors would like to thank Yoshiya Uwatoko for collaborative support. This work was partly supported by the National Natural Science Foundation of China (Grant No. 12204487, No. 12374136, No. 12274412, and No. U1932217), the National Key R$\&$D Program of China (Grant No. 2018YFA0704300 and No. 2021YFA1600201), the Strategic Priority Research Program (B) of the Chinese Academy of Sciences (Grant No. XDB25000000), the Fundamental Research Funds for the Central Universities 2242024k30029.

N. Z. and Y. S. contributed equally to this paper.

\bibliographystyle{apsrev4-2}
\bibliography{EuFeAsP}

\pagebreak
\newpage
\onecolumngrid
\begin{center}
	\textbf{\huge Supplemental information}
\end{center}
\vspace{1cm}
\onecolumngrid
\setcounter{equation}{0}
\setcounter{figure}{0}
\setcounter{table}{0}

\makeatletter
\renewcommand{\theequation}{S\arabic{equation}}
\renewcommand{\thefigure}{S\arabic{figure}}

\begin{figure}[h]
\includegraphics[width=17cm]{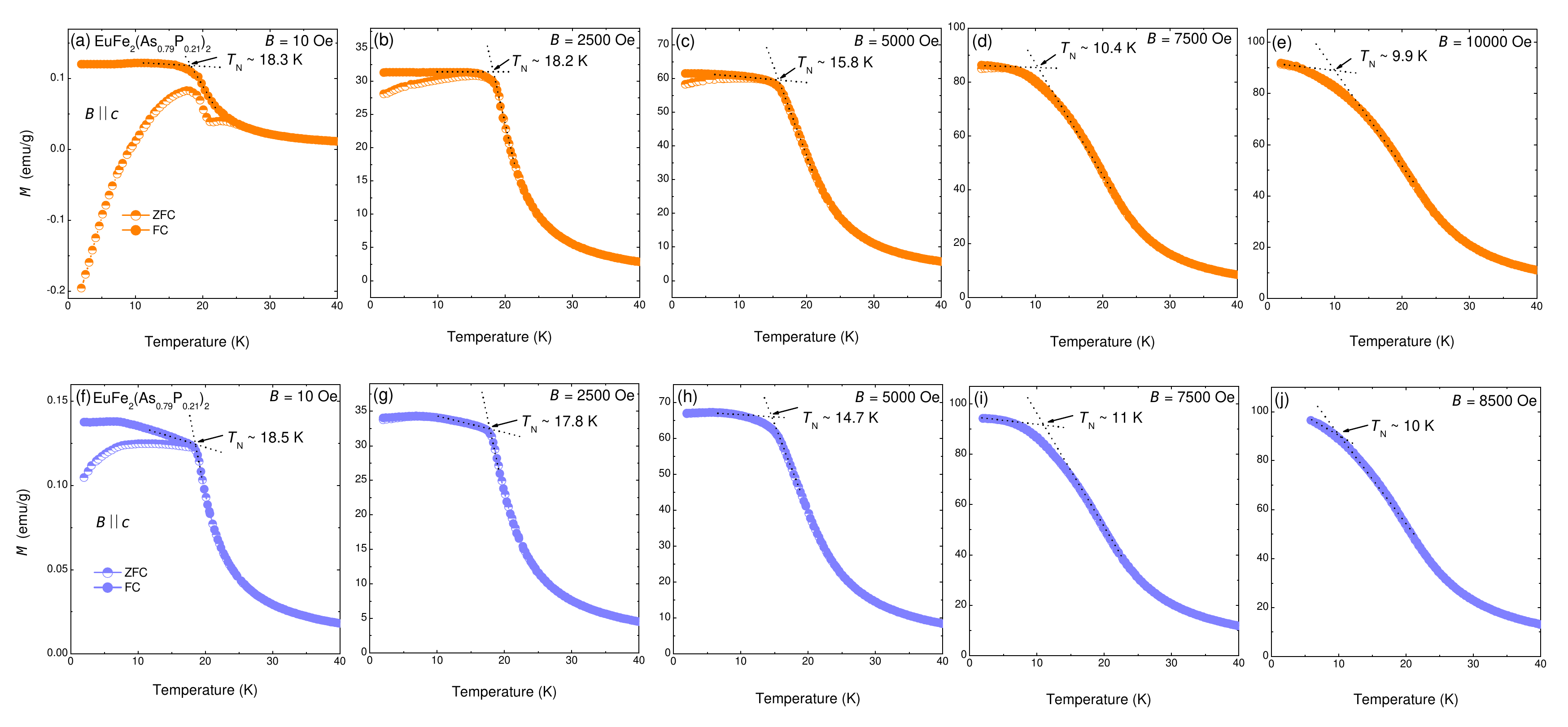}\\
\caption{\label{S2} The magnetizations with ZFC (half solid circle) and FC (solid circle) modes for (a)--(e) EuFe$_{2}$(As$_{0.79}$P$_{0.21}$)$_{2}$ and (e)--(j) EuFe$_{2}$(As$_{0.71}$P$_{0.29}$)$_{2}$, respectively.}\label{}
\end{figure}

\begin{figure}\center
\includegraphics[width=17cm]{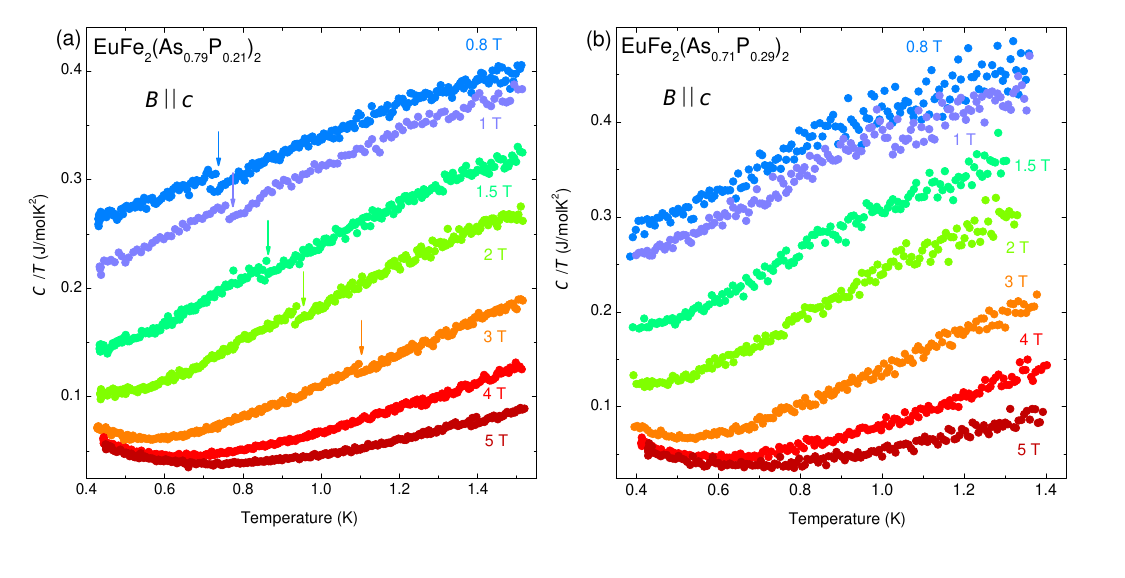}\\
\caption{\label{S3} (a) Temperature dependences of \textit{C}(\textit{T})/\textit{T} under various fields for EuFe$_{2}$(As$_{0.79}$P$_{0.21}$)$_{2}$ crystal (\textit{B} $\parallel$ \textit{c}). (b) Temperature dependences of \textit{C}(\textit{T})/\textit{T} under various fields for EuFe$_{2}$(As$_{0.71}$P$_{0.29}$)$_{2}$ crystal (\textit{B} $\parallel$ \textit{c}).}\label{}
\end{figure}

\subsection*{S1 The evolution of the out of and in plane angular dependent heat capacity}

\begin{figure}\center
\includegraphics[width=12cm]{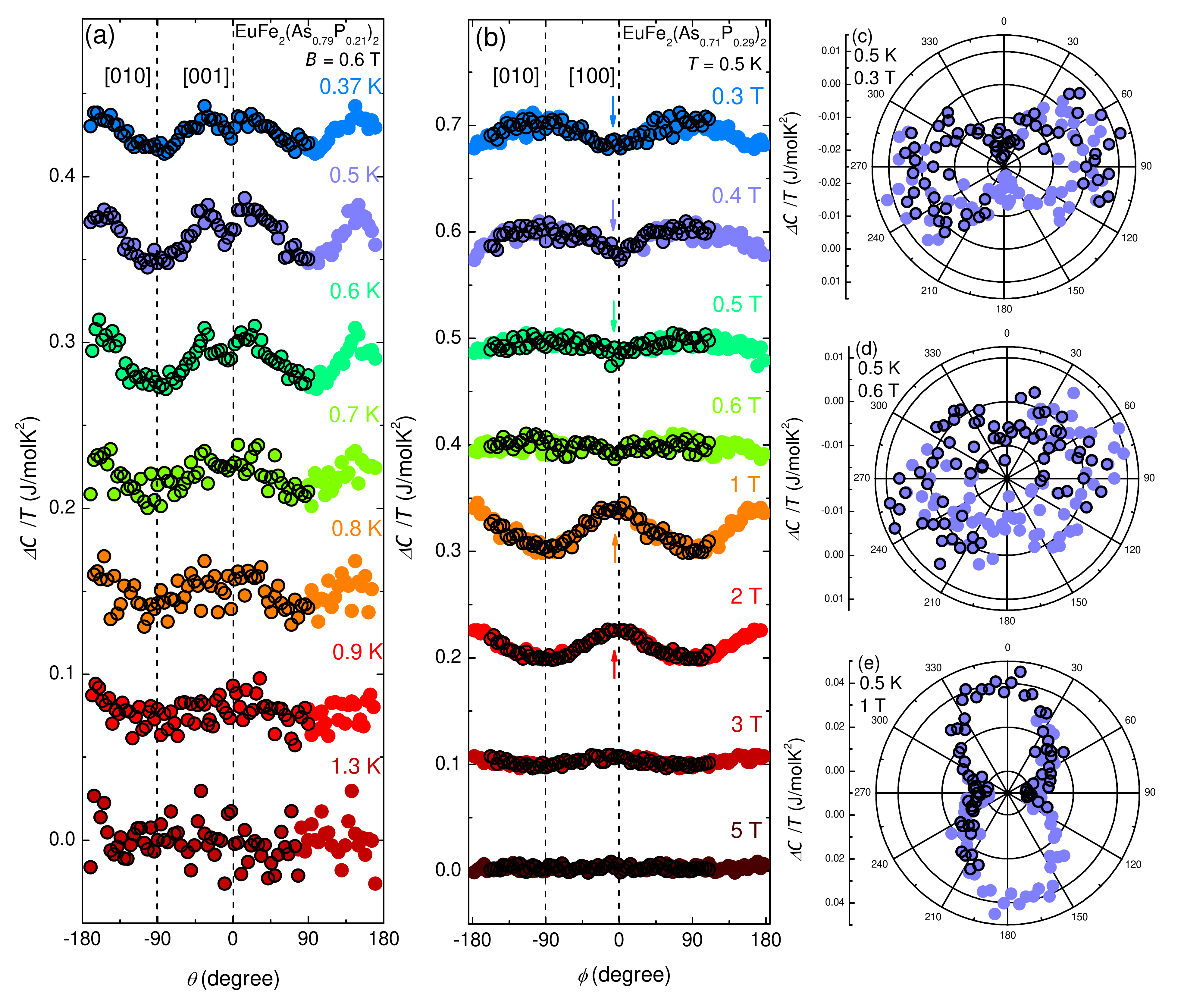}\\
\caption{\label{S4} (a) \textit{$\Delta$}\textit{C}(\textit{$\theta$})/\textit{T} measured under 0.6 T at various temperatures for EuFe$_{2}$(As$_{0.79}$P$_{0.21}$)$_{2}$ crystal. Each subsequent curve is shifted by 0.07 J/molK$^{2}$. (b) Azimuthal angle dependence of the specific heat \textit{$\Delta$}\textit{C}(\textit{$\phi$})/\textit{T} measured at 0.5 K under various fields for EuFe$_{2}$(As$_{0.71}$P$_{0.29}$)$_{2}$ crystal. Each subsequent curve is shifted vertically by 0.1 J/mol K$^{2}$. (c)--(e) Polar plot of the in-plane \textit{$\Delta$}\textit{C}/\textit{T} for EuFe$_{2}$(As$_{0.71}$P$_{0.29}$)$_{2}$ crystal at 0.5 K under 0.3, 0.6, and 1 T, respectively.}\label{}
\end{figure}

\begin{figure}\center
\includegraphics[width=15cm]{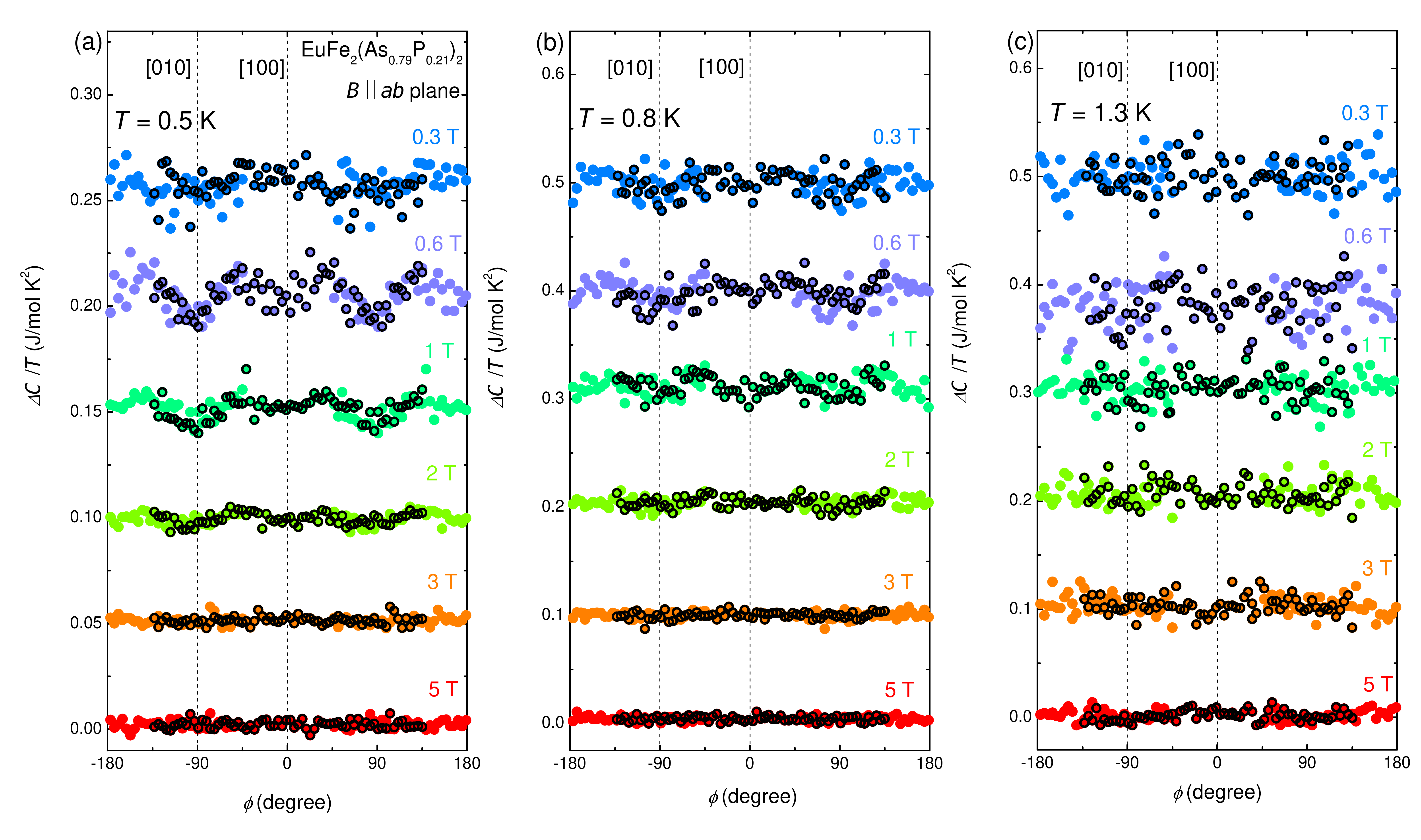}\\
\caption{\label{S5} Azimuthal angle dependence of the in-plane specific heat \textit{$\Delta$}\textit{C}(\textit{$\phi$})/\textit{T} for EuFe$_{2}$(As$_{0.79}$P$_{0.21}$)$_{2}$ crystal under various fields at (a) \textit{T} = 0.5 K, (b) \textit{T} = 0.8 K, and (c) \textit{T} = 1.3 K, respectively. \textit{$\Delta$}\textit{C}(\textit{$\theta$})/\textit{T} is defined as \textit{C}(\textit{$\theta$})/\textit{T}-\textit{C}(90$^{\circ}$)/\textit{T}, each subsequent curve is shifted by (a) 0.05 J/molK$^{2}$, (b) 0.1 J/molK$^{2}$, and (c) 0.1 J/molK$^{2}$.}\label{}
\end{figure}

\begin{figure}\center
\includegraphics[width=17cm]{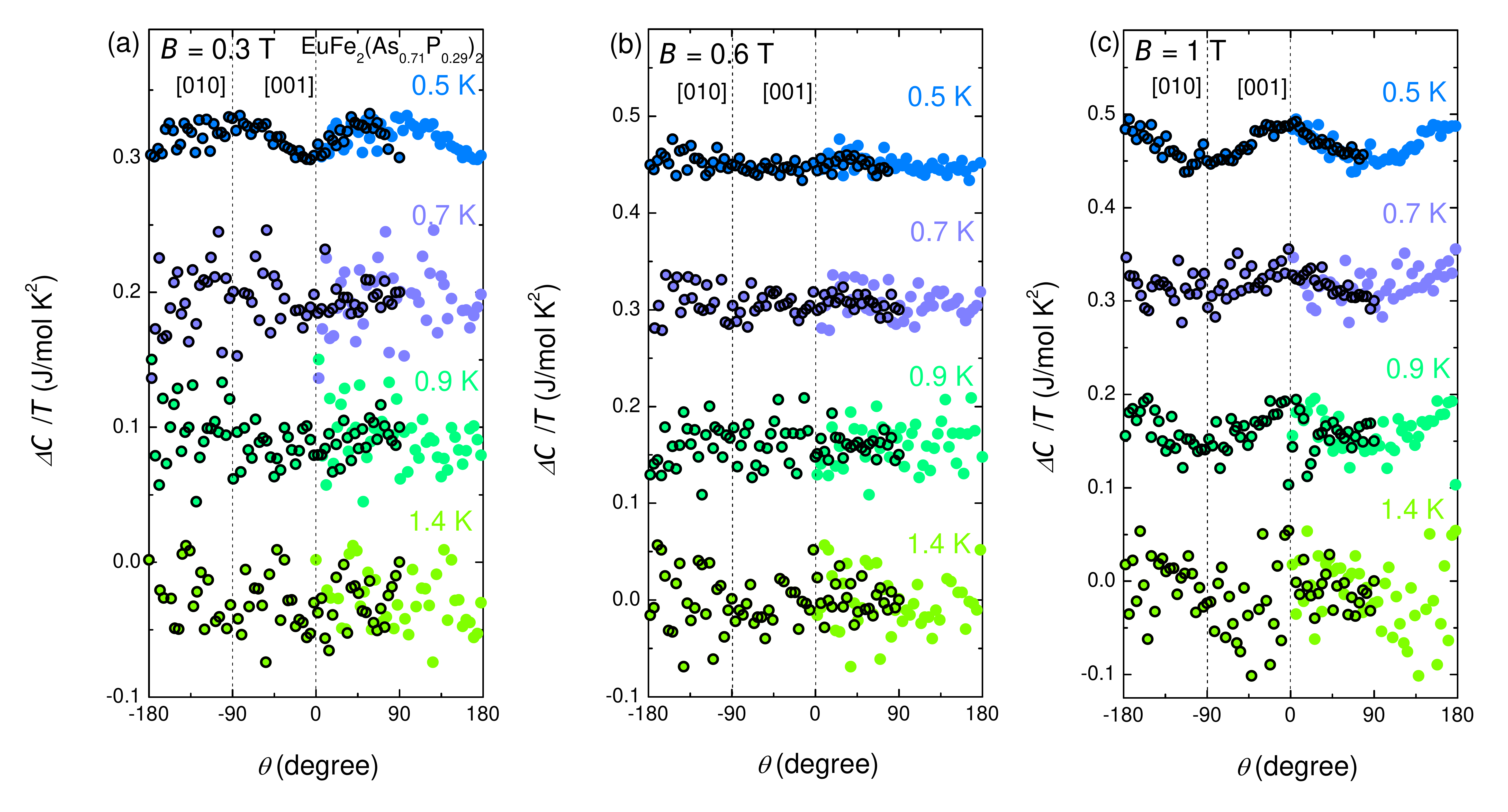}\\
\caption{\label{S6} Azimuthal angle dependence of the out of plane specific heat \textit{$\Delta$}\textit{C}(\textit{$\theta$})/\textit{T} for EuFe$_{2}$(As$_{0.71}$P$_{0.29}$)$_{2}$ crystal under various temperatures at (a) \textit{B} = 0.3 T, (b) \textit{B} = 0.6 T, and (c) \textit{B} = 1 T, respectively. \textit{$\Delta$}\textit{C}(\textit{$\theta$})/\textit{T} is defined as \textit{C}(\textit{$\theta$})/\textit{T}-\textit{C}(90$^{\circ}$)/\textit{T}, each subsequent curve is shifted by (a) 0.1 J/molK$^{2}$, (b) 0.15 J/molK$^{2}$, and (c) 0.15 J/molK$^{2}$.}\label{}
\end{figure}

More details about the evolution of the angular dependent specific heat under various fields and temperatures are shown in Figs. S3--S5. Figure S3(a) shows $\Delta$\textit{C}($\theta$)/\textit{T} under 0.6 T at various temperatures for the optimally doped crystal. Figures S3(b) shows $\Delta$\textit{C}($\phi$)/\textit{T} at 0.5 K under various fields for the overdoped crystal. Figures S3 (c)--(e) show the polar plot of the in-plane angular dependent heat capacity for the overdoped crystal. Figure S4 shows the angular dependence of specific heat under different fields and temperatures when the field is rotated in the \textit{ab} plane of EuFe$_{2}$(As$_{0.79}$P$_{0.21}$)$_{2}$. Figure S5 shows the angular dependent heat capacity data under various fields and temperatures when the field is rotated out of the \textit{ab} plane of EuFe$_{2}$(As$_{0.71}$P$_{0.29}$)$_{2}$. The ordinate \textit{$\Delta$}\textit{C}(\textit{$\theta$})/\textit{T} (out of plane) and \textit{$\Delta$}\textit{C}(\textit{$\phi$})/\textit{T} (in plane) are defined as \textit{C}(\textit{$\theta$})/\textit{T}-\textit{C}(90$^{\circ}$)/\textit{T} and \textit{C}(\textit{$\phi$})/\textit{T}-\textit{C}(90$^{\circ}$)/\textit{T}, respectively. In order to obtain a clear view of the symmetry, each subsequent curve is shifted vertically by a fixed value in Figs. S3, S4, and S5. Symbols with black outlines are measured data, and those without are mirrored points to show the symmetry. Definitions of azimuthal (\textit{$\phi$}) and polar (\textit{$\theta$}) angles with respect to the crystal axes are shown in fig. 3(a). These results show that the twofold symmetric curve could evolve to a four-peak heat capacity signal in the optimal crystal (\textit{x} = 0.21) as shown in Figs. 3 and S3(a). Meanwhile, it is found that the peak of the twofold symmetric curve could change to the valley by tuning the field in the overdoped crystal (\textit{x} = 0.29). It is an interesting phenomenon, which is worth studying in the future and will provide important insights into the understanding of these EuFe$_{2}$As$_{2}$-based materials.

\subsection*{S2 Dc magnetization measurements at very low temperatures for the P-doped crystals}

\begin{figure}\center
\includegraphics[width=17cm]{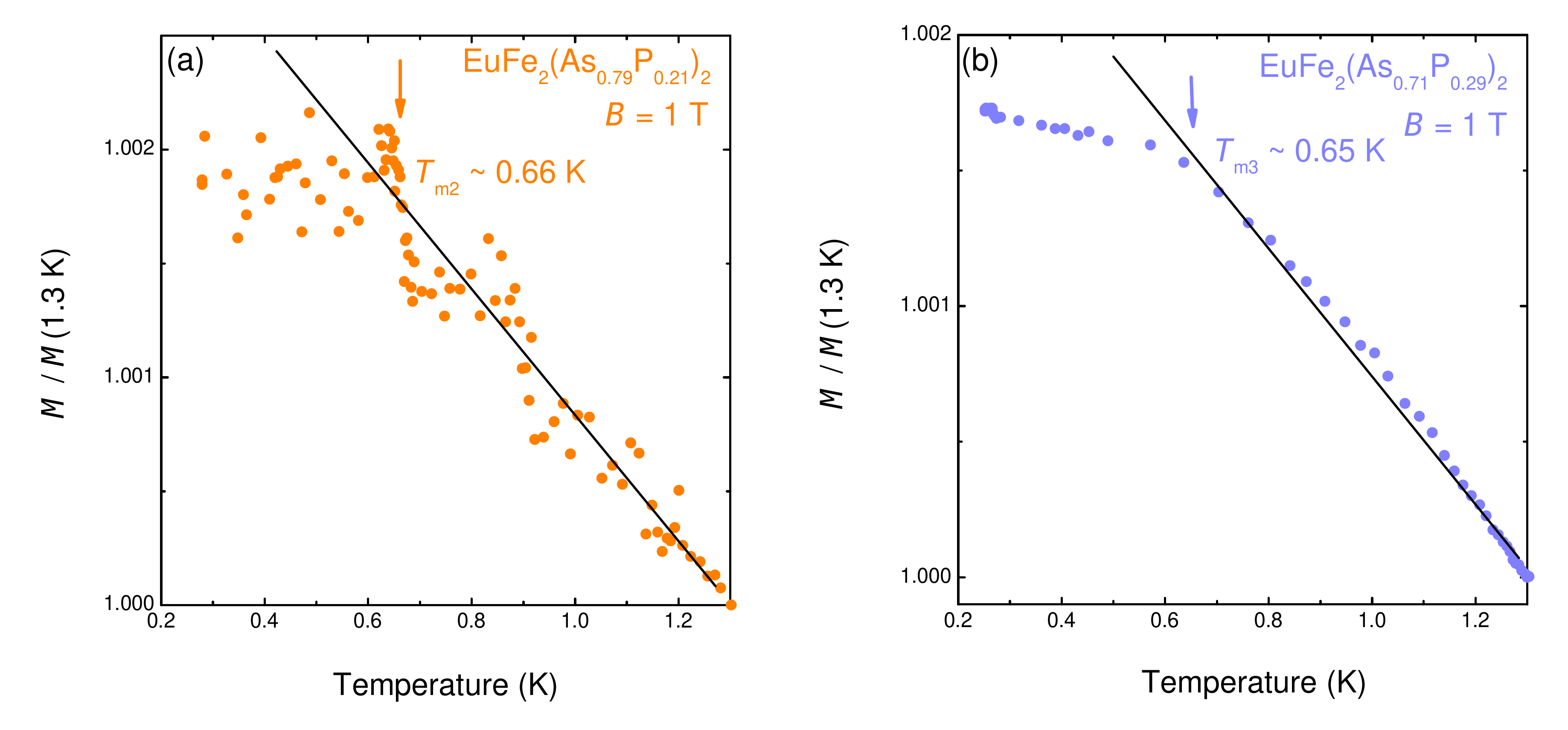}\\
\caption{\label{S8} Temperature dependent dc magnetization under 1 T for (a) the optimally doped crystals and (b) the overdoped crystals, which were measured by using the Capacitance-Faraday method.
}\label{}
\end{figure}

\begin{figure}\center
\includegraphics[width=17cm]{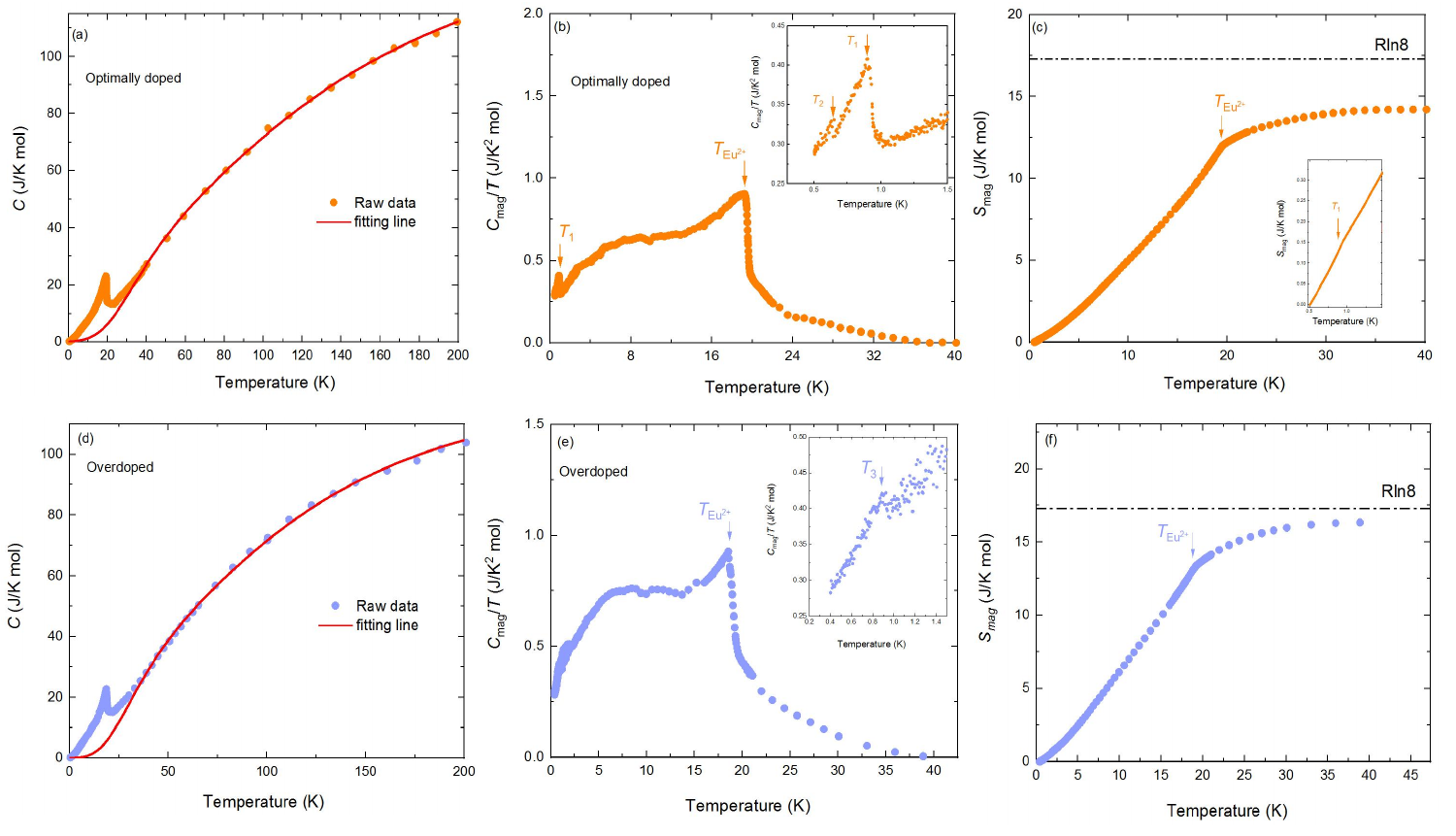}\\
\caption{\label{S7} (a) and (d) Heat capacity of the optimally and over doped crystal versus temperature from 0.4 to 200 K measured in \textit{B} = 0 T. The solid curve is a fit of the data by the Debye-Einstein model. The fitting could be divided into two steps. For the high temperature region (1.5 K to 200 K), we fit the data according to the equation of \textit{C}$_{p}$(\textit{T})=\textit{m}\textit{C}$_{D}$(\textit{T})+(1-$\textit{m}$)\textit{C}$_{E}$(\textit{T})+$\gamma$$\textit{T}$, \textit{C}$_{D}$(\textit{T})=9\textit{nR}($\frac{\textit{T}}{\textit{T}_D}$)$^{3}$$\int$$\frac{\textit{x}^{4}}{(e^{x}-1)(1-e^{-x})}$\textit{dx}, \textit{C}$_{E}$(\textit{T})=3\textit{nR}($\frac{\textit{T}_E}{\textit{T}}$)$^{2}$$\frac{1}{(e^{T_E/T}-1)(1-e^{-T_E/T})}$. For the low temperature region (0.4 K to 1.5 K), we fit the data according to the equation of \textit{C}$_{p}$(\textit{T})=$\gamma$\textit{T}+$\beta$\textit{T}$^{3}$. (b) and (e) Magnetic contribution to the heat capacity \textit{C}$_{mag}$ plotted as \textit{C}$_{mag}$(\textit{T})/\textit{T} versus \textit{T} for the two crystals, respectively. (c) and (f) Magnetic contribution to the entropy $\textit{S}$$_{mag}$($\textit{T}$).}
\label{}
\end{figure}

In order to identify if these low-temperature phase transitions have a relationship with the magnetism, we perform the dc magnetization measurements by using the Capacitance-Faraday method. When a specimen of magnetization \textbf{\textit{M}} is placed in an inhomogeneous magnetic field \textbf{\textit{H}}, it will experience a force \textbf{\textit{F}} = (\textbf{\textit{M}} $\cdot$ \textbf{\textit{$\bigtriangledown$}})\textbf{\textit{H}}. We could obtain the magnetization \textbf{\textit{M}} of the specimen by measuring this force, which has been known as Faraday method. The inhomogeneous magnetic field \textbf{\textit{H}} is generated by a specially-designed superconducting magnet that has gradient coils in addition to a main solenoid coil. And the force \textbf{\textit{F}} is measured by a transducer made of a parallel-plate capacitor.

The data were taken while cooling down with the applied field of 0.3 T (see the inset of Fig. 2(a) and 2(b)) and 1 T (see Fig. S6) for the two P-doped crystals, respectively.
Some obvious anomalous magnetizations have been observed in both the two P-doped crystals. At very low temperature range, the magnetization shows a rapid increase and lower temperature plateau, which could be understood to be related to a kind of ferromagnetic nature. Interestingly, the temperature dependent magnetization is almost consistent with the heat capacity behaviour. These results indicate that the origin of the low-temperature phase transitions has a close relationship with the magnetization, and the localized Eu$^{2+}$ spin order or their spin reorientation could serve as a possible candidate for understanding this phenomenon in EuFe$_{2}$As$_{2}$-based system.

\end{document}